\documentclass[dvips]{scrartcl}
\usepackage{amsmath}
\usepackage{slashed}
\usepackage{verbatim}
\usepackage{graphicx,epsfig}
\usepackage{hyperref}
\def\lsim{\mbox{\raisebox{-.6ex}{~$\stackrel{<}{\sim}$~}}}

\def\beq{\begin{equation}}
\def\eeq{\end{equation}}
\def\beqa{\begin{eqnarray}}
\def\eeqa{\end{eqnarray}}

\def\sfrac#1#2{{\textstyle{#1\over #2}}}

\title{Light dark matter\\
versus astrophysical constraints}
\author{    James M.\ Cline$^1$ \thanks{jcline@physics.mcgill.ca},
 Andrew R.\ Frey$^2$ \thanks{a.frey@uwinnipeg.ca}
  \\
  \textit{\large $^1$ Physics Department, McGill University, Montreal, QC, H3A2T8,
    Canada}\\
\textit{\large $^2$Dept.\ of Physics and Winnipeg Insitute for Theoretical 
Physics,}\\ \textit{\large The
University of Winnipeg, Winnipeg, MB, R3B2E9, Canada}	
}
\date{}

\begin{document}
\maketitle

\begin{abstract} 

Hints of direct dark matter detection coming from the DAMA, CoGeNT
experiments point toward light dark matter with isospin-violating and
possibly inelastic couplings.   However an array of astrophysical
constraints are rapidly closing the window on light dark matter. We
point out that if the relic density is determined by annihilation into
invisible states, these constraints can be evaded.  As an example we 
present a model of quasi-Dirac dark
matter, interacting via two U(1) gauge bosons, one of which couples to
baryon number and the other which kinetically mixes with the photon.
Annihilation is primarily into ``dark neutrinos'' that do not mix with
the SM, but which could provide an extra component of dark radiation. 
The model could soon be tested by several experiments searching for
such light gauge bosons, and we predict that both could be detected.
The model also requires a
fourth generation of quarks, whose existence might increase the
production cross section of Higgs bosons at the Tevatron and LHC.

\end{abstract}

{\bf Introduction.} 
The DAMA \cite{Bernabei:2008yi} and CoGeNT  
\cite{Aalseth:2010vx,Aalseth:2011wp}
experiments have presented evidence for light,
$\sim 10$ GeV dark matter (DM), which is at odds with null results from
Xenon10 \cite{Angle:2009xb}, Xenon100 \cite{Aprile:2011hi} and CDMS 
\cite{Ahmed:2009zw} for the simplest DM models, and moreover
the cross sections needed by DAMA and CoGeNT are at odds with each
other.  It is very intriguing that a single hypothesis, that DM
has isospin-violating interactions with nucleons, resolves the
discrepancies between all of the experiments except CDMS 
\cite{Chang:2010yk}-\cite{Gao:2011ka}.  One further
assumption, that the interactions are inelastic, connecting two
DM states split by $\sim 10$ keV, helps to alleviate the tension with 
CDMS \cite{Chang:2010yk,Frandsen:2011ts} (see also
\cite{Fox:2011px}).\footnote{The CRESST experiment \cite{Angloher:2011uu} reports evidence for light
DM with mass 9 GeV at the lower end of their M2 best fit region
and cross section $\sigma_n\sim 1.5\times 10^{-4}$ pb.  In an isospin violating 
model with $f_p/f_n = -1.5$, this value would increase by a factor of
$[\frac12(1+f_p/f_n)]^{-2} = 16$, since O and Ca have equal numbers of
protons and neutrons, which is a factor of 13 below the inelastic CoGeNT/DAMA
best-fit value for $\sigma_n = 0.03$ pb which we adopt below \cite{Farina:2011pw}.  
This discrepancy is greatly reduced (to a factor of 2) if one adopts 
the value $\sigma_n = 5\times 10^{-3}$ pb, corresponding to the best
fit with elastic scattering.}

Recently we proposed some ``minimal models'' of hidden sector dark 
(DM) matter \cite{Cline:2011zr} that have the desired properties, and
we noted that a number of astrophysical considerations constrain the
models very strongly.  Essentially, if DM of mass $\sim 10$ GeV
annihilates primarily into any channel other than muons, it is ruled
out by constraints from dumping electromagnetic energy into the cosmic
microwave background (CMB)  \cite{Hutsi:2011vx,Galli:2011rz}, Fermi
observations of dwarf satellite galaxies
\cite{GeringerSameth:2011iw,collaboration:2011wa}, or  SuperK limits
on neutrinos from DM annihilations in the sun
\cite{Kappl:2011kz,Chen:2011vd,Gao:2011bq}.   These constraints will tighten as a
result of forthcoming data from new experiments like Planck.  Moreover
the PAMELA constraint on cosmic ray antiprotons excludes 10 GeV DM
with an annihilation cross  section greater than 0.1 times the
standard relic density value if the final state contains quarks that
can hadronize  \cite{Cao:2009uv}.  This tension is demonstrated to be
rather insensitive to different choices for cosmic ray propagation
models and halo models for the $b\bar b$ channel in
\cite{Lavalle:2010yw}.  The only robust particle physics mechanism for
evading this tension is if the annihilation is into a pair of bosons
that are too light to decay into $p$-$\bar p$ (or do not couple to
quarks).

Of course one of these new experiments may find deviations giving
positive indirect evidence for light dark matter.  But if they instead
only tighten the constraints, while direct evidence for light DM
persists, we will have a puzzle.  One elegant way out is asymmetric
dark matter that carries a conserved charge.  If the symmetric
component of the DM has annihilated away, then the above constraints
no longer apply.  In the present letter, we wish to point out a
different possibility, assuming the DM is symmetric.  Namely, if it
annihilates primarily into invisible particles, the constraints in
question are evaded.  While this may seem like a trivial statement,
in fact care must be taken to avoid problems from other constraints,
such as the number of species during big bang nucleosynthesis
(BBN) or too-rapid cooling of supernovae by emission of the new
invisible states.\footnote{We thank D.\ Spolyar for reminding us about
this important constraint.}  

To make these points concrete we will illustrate them in a hidden
sector model similar to one 
presented in \cite{Cline:2011zr}.  This model achieves isospin
violation using a light vector boson $B_\mu$ that couples
both to gauged baryon number and, and another one
$Z'$ that kinetically mixes with the photon.
The DM gets inelastic couplings and a
small  mass splitting through a weak Yukawa interaction with a new
Higgs field that breaks the U(1) gauge symmetries.   It annihilates
dominantly into light ``dark neutrinos'' via exchange of another new
Higgs field, which is also the one primarily responsible for the $\sim
10$ GeV DM mass.   Thus the model is economical, in that most of
elements serve for more than one purpose.   An interesting feature of
this model is that the kinetic mixing of the $Z'$ boson is related
to its mass in such a way that it might be discovered in the near future
by several proposed experiments aimed at detecting such particles. 
Moreover because the $B$ boson might mix weakly with $Z'$, it could also be
found in the same searches.

{\bf The Model.} Gauged baryon ($B$) number is anomalous in the standard model,
but the anomalies can be canceled by adding a fourth generation of  quarks with
baryon number $\pm 1$  \cite{Carone:1994aa,FileviezPerez:2011pt}.  If we wish to
further couple  the DM $\chi$ to the associated vector boson $B_\mu$, then to
avoid  further anomalies it is natural to assume that  $\chi$ is initially
vector-like (Dirac), but a new Higgs boson $\phi$ that spontaneously breaks some
combination of the dark U(1) symmetrires can also couple to $\chi$ and render it
quasi-Dirac after symmetry breaking.  This fits in nicely with the additional 
preference that  the DM interactions with nuclei should be inelastic. A small
$\sim 10$ keV mass splitting between the two DM components is thus both welcome
and natural. But because the preference for inelasticity is weak in global fits
to all data \cite{Farina:2011pw}, we will consider two versions of the model,
one elastic and the other inelastic. To get isospin violating couplings of the
DM to nucleons, we give the DM an additional U(1) coupling to a vector $Z'$ that
kinetically mixes with the photon.

The dark sector field content consists of two Weyl DM components 
$\chi_{1,2}$ with
charges $\pm(g_B,g_{Z'})$ under U(1)$_B\times$U(1)$_{Z'}$,
a real singlet $\Phi$,  
a complex Higgs $\phi'$ with charge $(0,g'_{Z'})$,
another $\tilde\phi$ with charge $(\tilde g_B$,0), and in the 
inelastic version of the model, a third $\phi$ with
charge $(-2g_B,-2g_{Z'})$   There
are also ``dark neutrinos'' $\nu_1$, $\nu_2$ that carry no gauge
quantum numbers.  All of the Higgs fields need to get VEVs in order to give
masses to the dark sector particles. The interactions are
\beqa
	V &=& y_\chi\Phi\chi_1\chi_2 + y_\nu\Phi\nu_1\nu_2 + M_\nu \nu_{2} \nu_{2} 
	  + \tilde g_B^2|\tilde\phi|^2  B^2 + {g'_{Z'}}^{\!\!\!2}|\phi'|^2 Z'^2\nonumber\\
	&+& (g_B B_\mu + g_{Z'} Z'_\mu)(\chi_1^\dagger \sigma^\mu \chi_1 - 
	\chi_2^\dagger \sigma^\mu \chi_2) +
	 g_B\bar N\slashed{B} N + \epsilon e\bar p \slashed{Z'} p\nonumber\\
	&+&\left\{\begin{array}{ll} 4(g_B B_\mu + g_{Z'} Z'_\mu)^2|\phi|^2+ 
	\sfrac12 y_\phi(\phi\chi_1\chi_1 + \phi^*\chi_2\chi_2),&
	{\rm inelastic\ model}\end{array}\right\}
\label{Veq}
\eeqa
where $p$ is the proton,
and $N$ is either kind of  nucleon.  Note that for the elastic version of the 
model, the last line is omitted, and expressions like $\chi_1\chi_2$ are
shorthand for $\chi_1^T\sigma_2\chi_2$.  The lighter dark  
neutrino which we will call $\nu\,'$ gets a small mass from the see-saw
mechanism once $\Phi$ gets a VEV, due to the large bare mass
$M_\nu$.  The structure of the couplings of
$\Phi$ can be justified by a discrete symmetry\footnote{
This symmetry must
be softly broken to avoid the cosmological  domain wall problem,
but such breaking in the potential for $\Phi$ does not concern
us in the following.} $\chi_1\to
-\chi_1$, $\nu_1\to -\nu_1$, $\Phi\to -\Phi$.
The light dark neutrino mass eigenstate $\nu\,'$ allows the DM to annihilate invisibly via
$\chi\chi\to\Phi\Phi$ followed by  $\Phi\to\nu\,'\nu\,'$.
For simplicity we have imposed a second discrete symmetry $\chi_1\to\chi_2$,
$\phi\to\phi^*$ so that there is only a single Yukawa coupling $y$.
When $\Phi$ and $\phi$ get their VEVs, the DM mass  eigenstates are $\chi_\pm = 
\sfrac{1}{\sqrt{2}}(\chi_1\pm\chi_2)$ with masses $M_\pm = M_\chi\pm\mu$
where $M_\chi = y_\chi\langle\Phi\rangle$ and $\mu = 
y_\phi\langle\phi\rangle$.  

The gauge bosons $B$ and $Z'$ do not mix with each other in the elastic version
of the model, but the VEV of $\phi$ causes them to do so in the  inelastic
version.   We want to suppress such mixing to avoid having any gauge boson with
significant couplings to both baryon and lepton number, since the constraints on
such particles are quite severe.  (These constraints also place stringent
limits on isospin violating dark matter models in which a single vector
couples to baryon number and mixes kinetically with the photon.)  
Ref.\ \cite{Williams:2011qb} shows that a
vector that couples to $B$ and has kinetic mixing of $\epsilon = 10^{-3}$ is
relatively safe, but one with $\epsilon = 10^{-2}$ is ruled out if its mass is
less than 10 GeV.  Moreover our determination of the correct amount of isospin
violation is simpler if the mixing between $B$ and $Z'$ is negligible so that we
can consider these fields to coincide with the mass eigenstates.  We will thus
assume that $\langle\phi\rangle$ is significantly smaller than
$\langle\tilde\phi\rangle$ and $\langle\phi'\rangle$. In the mass eigenstate
basis, the interaction of $\chi$ with $B$ becomes purely off-diagonal:
$g_B\bar\chi_+\slashed{B}\chi_- +$ {\rm h.c.}, written in terms of
Majorana-Dirac spinors.

{\bf Fitting to CoGeNT/DAMA and relic density.} 
We now consider the constraints arising from direct and indirect
detection. (See table
\ref{tab1} for a summary of the results for constraints on all the
model parameters.)
Comparing the diagrams for $\chi$-$n$ and $\chi$-$p$ scattering,
the ratio of DM couplings to neutrons and protons is 
\beq
	{f_p\over f_n} = 1 + {g_{Z'}\epsilon e\over g_B^2}\,{m^2_B\over
	m^2_{Z'}}
\label{fpfn}
\eeq
We fix $f_p/f_n= -1.53$, as needed to evade bounds from Xenon
and to reconcile the DAMA/CoGeNT observations \cite{Farina:2011pw}. 
The overall rates
for DAMA/CoGeNT are reproduced by matching the cross section for
scattering on neutrons, $\sigma_n  = g_B^4\mu_n^2/(\pi m_B^4)$
to the value $3\times 10^{-38}$ cm$^2$ \cite{Farina:2011pw,Cline:2011zr}, where the reduced
nucleon mass is $\mu_n = 0.84$ GeV if the DM mass is $M_\chi=8$ GeV.
This gives $m_B/g_B = 232.3$ GeV and $m_{Z'}^2/(g_{Z'}\epsilon) = 
-(79.9{\rm\ GeV})^2$.   

\begin{figure}[htb]
\centering
\includegraphics[width=0.8\textwidth]{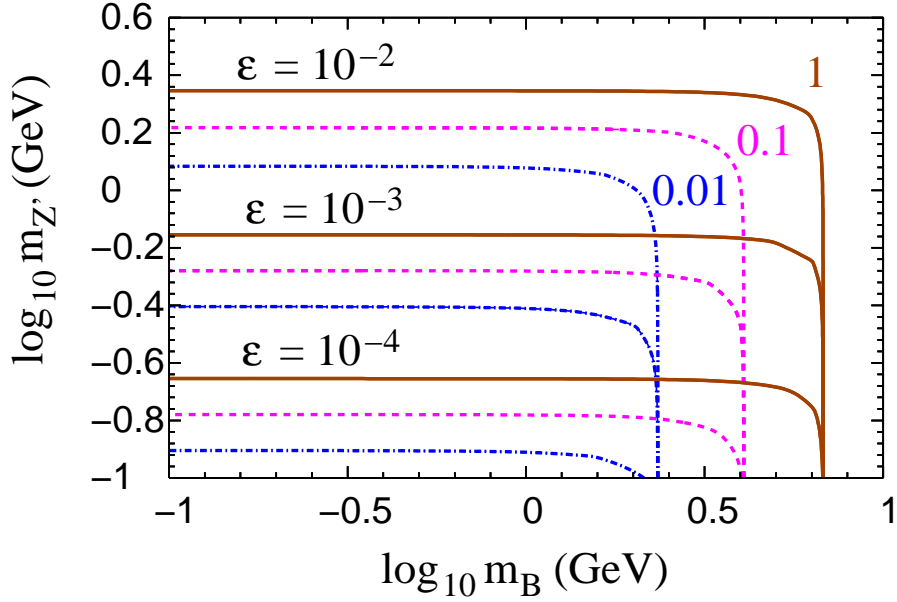}
\caption{\label{sigann} For kinetic mixing $\epsilon = 10^{-2},
10^{-3},10^{-4}$, we show contours in the $m_B$-$m_{Z'}$ plane
where the DM annihilation cross section into visible particles is
equal to 1 (solid), 0.1 (dashed) and 0.01 (dot-dashed) times the standard relic density value
1 pb$\cdot c$.  The allowed regions are below and to the left of the 
corresponding contour.}
\end{figure}

As described in \cite{Cline:2011zr}, we can use the above
relations between $\{m_B,g_B\}$ and $\{m_{Z'},g_{Z'}\}$ to eliminate the couplings in terms of the masses
when computing the annihilation cross section into
standard model particles by the processes $\chi\chi\to BB,Z'Z',BZ'
\to ff\bar f\bar f$ or $\chi\chi\to B,Z'\to f\bar f$.  For a given
choice of $\epsilon$, this results in contours of $\langle\sigma_{\rm
ann} v\rangle$ in the $m_B$-$m_{Z'}$ plane.  By comparing the
resulting value of $\langle\sigma_{\rm ann} v\rangle$ to that required
for the standard relic density, we see which ranges of the masses
are compatible with annihilation that is subdominantly into visible
particles, so that the relic density can be determined by invisible
annihilations.  The result is given in fig.\
\ref{sigann}, which  shows that for a given value of $\epsilon$, there
exists a region where $m_{Z'}$ and $m_{B}$ are both  bounded from above
such that annihilation into SM particles
is sufficiently suppressed.  For a given value of $\epsilon$, we find that the
maximum allowed value of $m_{Z'}$ is given by 
\beq
	m_{Z'} < \sqrt{\epsilon}\cdot 16.6{\rm\ GeV}
\label{mzbound}
\eeq
which corresponds to the horizontal part of the contours showing where
$\langle\sigma_{\rm ann} v\rangle < 0.1\cdot
\langle\sigma_{\rm ann} v\rangle_0$ in fig.\ \ref{sigann}. This is the minimum reduction 
of annihilation into visible particles needed to satisfy the
astrophysical constraints.  On the other hand, $m_B < 4$ GeV is the bound on
$m_B$ regardless of $\epsilon$.

To get the correct relic density, it is convenient to assume that 
$m_{\Phi} < M_\chi$ so that $\chi\chi\to\Phi\Phi$ is
allowed.\footnote{If $m_{\Phi}> M_\chi$, one must compute the
annihilation cross section for $\chi\chi\to 4\nu\,'$.} In this case we
can determine a function of $y_\chi$ and $m_{\Phi}$ by  demanding that
the cross section for $\chi\chi\to\Phi\Phi$ be  $3\times 10^{-26}$
cm$^3$/s.  The theoretical cross section is given by
$\langle\sigma_{\rm ann} v\rangle = y_\chi^4 f(m_{\Phi}/M_\chi)/
(64\pi M_\chi^2)$, where $f(x) = (1-x^2)^{3/2}/(1-x^2/2)^2$.  For
$m_{\Phi}<0.9\, m_\chi$, $y_\chi$ is in the range $0.076$ to $0.11$.  
This implies that $\langle\Phi\rangle \sim 73-105$ GeV, since 
$M_\chi=y_\chi\langle\Phi\rangle \cong 8$ GeV is determined by the fit
to CoGeNT/DAMA \cite{Farina:2011pw}.  We find that the
$s$-channel contribution to $\chi\chi\to\Phi\Phi$, which is controlled by
the cubic term in potential for $\Phi$, is smaller than the dominant
$t$-channel contribution by a factor of
$m_{\Phi}^4/(16 y_\chi^2 M_\chi^2\langle\Phi\rangle^2)\ll 1$ in the
squared amplitude.   Furthermore, annihilation to
$\nu\,' \nu\,'$ is suppressed because, as we show below, the Yukawa
coupling $y_\nu$ is much smaller than $y_\chi$.

{\bf New Higgs boson parameters.}
Next we turn to properties of the other new Higgs bosons and 
Yukawa couplings.   For the inelastic version of the model, we
require that $g_B g_{Z'}\langle\phi\rangle^2\ll m_B^2$, $m_{Z'}^2$ to suppress
the gauge boson mixing.  If the gauge couplings are of similar size, this
implies $\langle\phi\rangle\ll \langle\tilde\phi\rangle$, $\langle\phi'\rangle$.
Since the mass splitting between the DM states is given by
$2\mu = 2 y_\phi\langle\phi\rangle\cong 10$
keV, having small $\langle\phi\rangle$ helps to explain the smallness
of this splitting.  For example if $\langle\phi\rangle\sim 10$ GeV,
then $y_\phi\sim 10^{-6}$.

The other VEVs $\langle\tilde\phi\rangle$, $\langle\phi'\rangle$,
can be be relatively large, if the couplings $\tilde g_B$, $g'_{Z'}$ are
small.  Using the constraints on $(m_B,g_B)$ and
$(m_{Z'},g_{Z'})$ found above, we have
\beq
	\langle\tilde\phi\rangle = {g_B\over \tilde g_B}\cdot 232{\rm\ GeV},
	\quad 
	\langle\phi'\rangle  = {\sqrt{|g_{Z'}\epsilon|}\over g'_{Z'}}
	\cdot 80{\rm\ GeV}
\label{phibounds}
\eeq
In particular, by choosing $\tilde g_B$ to be smaller than $g_B$,
$\tilde\phi$ can have a larger VEV than the SM Higgs.  This
helps the model to avoid being marginalized by recent very stringent
constraints on the fourth generation of quarks needed for cancellation
of the anomaly of gauged baryon number, since 
$\langle\tilde\phi\rangle$ can contribute to the mass to vector-like
4th generation quarks \cite{FileviezPerez:2011pt}. We discuss this
issue further in the final section.  

In the potential (\ref{Veq}) we have omitted the renormalizable
couplings of $\phi$ and $\Phi$ to the standard model Higgs
field,\footnote{$\phi'$ of course also has such couplings but is 
unimportant to our discussion because it does not couple to the DM.}
\beq
	\lambda_{\phi h} |\phi|^2 h^2 + 
	\lambda_{\Phi h} |\Phi|^2 h^2
\eeq
which would give rise to $\phi$-$h$ and $\Phi$-$h$ mixing,
with the mixing angles
\beq
	\theta \cong {\lambda_{\phi h}\langle\phi\rangle v\over
	m^2_h - m^2_{\phi}},\quad
	\Theta = {\lambda_{\Phi h}\langle\Phi\rangle v\over
	m^2_h - m^2_{\Phi}},\quad
\eeq
where $v=\langle h\rangle = 246$ GeV.  Such mixing gives rise to additional
interactions between the DM and nucleons by exchange of the new Higgs
particles.  Because $y_\phi$ is so small, there is no
strong constraint on $\theta$, but $\Phi$ exchange is potentially
dangerous since it would perturb  the amount of isospin
violation away from the optimal value if its effect on $\chi$-nucleon scattering was comparable to
that of the gauge boson exchange.  Moreover the inelastic feature would be
spoiled by strong $\Phi$ exchange since the couplings of
$\Phi$ are not off-diagonal in the $\chi$ mass eigenbasis.
To avoid this problem we require that 
\beq
	{y_\chi \Theta\, y_n\over m_{\Phi}^2}
	\ll {g_B^2\over m_B^2}
\eeq
where $y_n \cong 0.3\, m_n/v$  is the coupling of $h$ to the nucleon
(see for example \cite{Barger:2010yn}). 
However using the above values for $g_B/m_B$ and $y_\chi$,
this is not very stringent; even for $m_{\Phi}= 4$ GeV it only
constrains $\Theta \ll 2.6$, which far weaker than direct
experimental constraints on the mixing of such a light Higgs boson,
$\Theta < 10^{-2}$ \cite{O'Connell:2006wi}. 

{\bf Constraints on dark neutrinos.}
Finally we come to the properties of the light dark neutrino, $\nu\,'$.
It was last produced when the temperature of the universe was 
$O(m_{\Phi})$ from the decays 
$\Phi\to\nu\,'\nu\,'$.  If $m_{\Phi}\sim$ 1 GeV, this occurred before the QCD phase
transition, and the relic density of $\nu\,'$ is suppressed by a factor of
$\sim 60$, the reduction in the number of degrees of freedom in the 
thermal plasma.  This means that cosmological upper bounds on the sum of neutrino
masses, which we take to be $0.3$ eV (see  \cite{Abazajian:2011dt} for
a review),  are relaxed by a factor of $\sim 22$ (using the fact that
ordinary
neutrinos have a number density that is $4/11$ that of photons), 
giving the bound $m_{\nu\,'} < 6.5$ eV.  This puts an upper limit on 
$y_\nu$, 
\beq
	y_\nu \lsim 3.5\times 10^{-5}\left(M_\nu\over 1{\rm\
TeV}\right)^{1/2}
\label{ynubound}
\eeq
Because of the dilution of the $\nu\,'$ number density, there is no
constraint from its contribution to the Hubble rate during
BBN.

A potentially serious constraint arises from the fact that $\nu\,'$ can
be produced in supernovae by the nucleon collisions $NN\to NN\nu\,'\nu\,'$ (via
virtual $\Phi$ emission), causing them to cool more quickly than
observed in SN 1987A.  A similar process in which sterile neutrinos
are produced by vector current couplings to nuclei was considered
in ref.\ \cite{Grifols:1997iy}.  (The vector coupling is expected to
give similar results to our scalar coupling \cite{Georg}.)
Adapting these results to the present
case, we find that 
\beq
	{\sqrt{\Theta\,y_n\,y_\nu}\over m_{\Phi}}
	\lsim {1\over 100{\rm\ TeV}}
\eeq 
Using our fiducial values $m_{\Phi}\sim 1$ GeV and
$y_\nu \sim 3.5\times 10^{-5}$, this translates into
the constraint $\Theta < 2\times 10^{-3}$, which puts the
mixing of $\Phi$ with $h$ out of experimental reach 
\cite{O'Connell:2006wi} unless 
$y_\nu$ is taken to be smaller.

For DM to annihilate invisibly,  the decay channel
$\Phi\to\nu\,'\nu\,'$  must dominate over decays into standard model
channels like $\mu^+\mu^-$\footnote{in general, the heaviest SM
fermions that can be produced by $\Phi$ decay, keeping in mind that
$m_\Phi < M_\chi \cong 8$ GeV}
  due to the mixing
$\Theta$, implying that $y_\nu^2 \gg 
(\Theta y_\mu)^2$
(where $y_\mu = 4.3\times 10^{-4}$ is the SM Yukawa coupling of 
$\mu$).  If $\Theta$ saturates its direct experimental upper bound of $10^{-2}$
and $y_\nu$ saturates (\ref{ynubound}) then this criterion is 
satisfied even for $M_\nu$ at the TeV scale.  Heavier decay products
can be accommodated by a modest increase in $M_\nu$.

Interestingly, positive evidence for an additional species of
dark neutrinos has come from recent CMB data \cite{darknu}.
If $m_\Phi$ is below the QCD scale, the decays $\Phi\nu\,'\nu\,'$
can occur sufficiently late for $\nu\,'$ to have as large an abundance
as the SM species.  Smaller values of $m_\Phi$ would naturally be
correlated with lower $\nu'$ masses since both are related to
$\langle\Phi\rangle$.   These neutrinos must be sufficiently light
to constitute an extra component of radiation before recombination,
rather than being an extra component of the dark matter.

\begin{table}[t]
\begin{center}
\begin{tabular}{|c|c|c|}
\hline
parameter & value  & constraint\\
\hline
$-\epsilon$ & $< 0.03$ & fig.\ \ref{fig:apex}\\
$m_B$ & $<4$ GeV& fig.\ \ref{sigann} \\
$m_{Z'}$ & $< \sqrt{\epsilon}\cdot 16.6$ GeV & eq.\ (\ref{mzbound})\\
$g_B$ & $m_B/(232{\rm\ GeV})$  & eq.\ (\ref{fpfn}), $\sigma_n$\\
$g_{Z'}$ & $-\epsilon^{-1} m_{Z'}^2/(80{\rm\ GeV})^2$ &  eq.\ (\ref{fpfn}), $\sigma_n$\\
$\langle\phi\rangle$ & $\ll  \langle\phi'\rangle,
\langle\tilde\phi\rangle$ &  gauge boson mixing  \\
$\langle\tilde\phi\rangle$ &  $(g_B/\tilde g_B)\cdot 232$ GeV& eq.\
(\ref{phibounds})\\ 
$\langle\phi'\rangle$ &  $(\sqrt{|g_{Z'}\epsilon}/g'_{Z'})\cdot 80$ GeV& eq.\
(\ref{phibounds})\\ 
$\langle\Phi\rangle$ & $73-105$ GeV & $M_\chi/y_\chi$ \\
$y_\chi$ & $0.076-0.11$ &  $M_\chi$, relic abundance\\
$y_\phi$ & $\sim 10^{-6}$ & $\chi$ mass splitting \\
$y_\nu$ & $\lsim 3.5\times 10^{-5}$ & $m_{\nu\,'}$, $\tilde\phi\to
\nu\,'\nu\,'$\\
$M_\nu$ & $\sim$ TeV & $m_{\nu\,'}$\\
\hline
\end{tabular}
\end{center}
\caption{\label{tab1} Summary of allowed values of parameters of the
model (\ref{Veq}), and the corresponding constraints which determine
them.}
\end{table}

\begin{figure}[htb]
\centering
\includegraphics[width=0.8\textwidth]{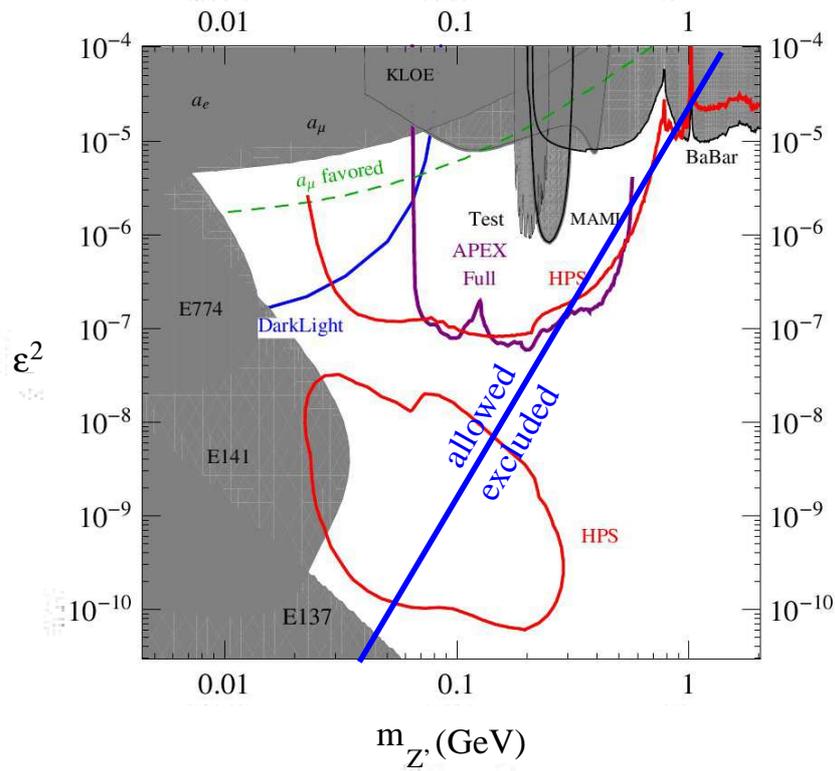}
\caption{\label{fig:apex} Solid (blue) line shows the maximum
allowed value of $m_{Z'}$ (eq.\  (\ref{mzbound})) in the dark matter 
model
for a given value of $\epsilon$, in the
$m_{Z'}$-$\epsilon^2$ plane; dark regions are
already excluded by various searches for light vector bosons, and
light circumscribed regions are targeted by upcoming experiments. 
 Background figure courtesy of R.\ Essig. 
}
\end{figure}

{\bf Discussion.}  We have presented what we believe to be the
simplest model of symmetric light dark matter that has the potential
for explaining the tentative evidence from CoGeNT and
DAMA\footnote{from footnote 1, it appears that 
that CRESST can only be made compatible with these if the requirement
of inelastic scattering is dropped} while robustly evading stringent
astrophysical constraints that may soon exclude all such models unless
the dominant DM annihilations are  into invisible particles.  The
isospin violating interactions  between DM and nucleons are mediated
by light vector bosons $B$ of gauged baryon number and a $Z'$ that has
kinetic mixing $\epsilon$ to the photon.   

The model is strongly constrained, and predicts a maximum value of
$m_{Z'}$ for each  value of $\epsilon$ (\ref{mzbound}) to sufficiently
suppress the visible annihilation channels, as well as a minimum value
of $m_B$ to suppress the gauge boson mixing angle.  There are shown in
fig.\ \ref{fig:apex}. The region of the $m_{Z'}$-$\epsilon$ plane to
the left of the solid line is  interesting from the point of view of the
proposed Heavy Photon Search (HPS) \cite{hps} and DarkLight
\cite{darklight,Freytsis:2009bh} experiments at Jefferson Laboratory.
Fig.\ \ref{fig:apex} shows that these two experiments, as well as the
already-running APEX experiment \cite{Abrahamyan:2011gv}, are capable
of probing a large fraction of the parameter space predicted by the
model.  We make the interesting observation that both $B$ and $Z'$
could be discovered by these experiments in the inelastic version of the 
model, where $B$ has a small mixing $\theta_B$ with $Z'$ and thus couples to 
the electromagnetic current with strength $\epsilon\theta_B$.

Another experimental test of the model will come from the requirement
of a fourth generation of quarks carrying exotic baryon number $\pm 1$
in order to cancel gauge anomalies of $B$ 
\cite{Carone:1994aa,FileviezPerez:2011pt}.  Because the field
$\tilde\phi$ that spontaneously breaks $B$ can have a larger VEV than
the SM  Higgs, it can increase the mass of vector-like exotic quarks
without requiring them to have very large Yukawa couplings. Otherwise,
such exotic quarks would be  difficult to hide from current collider
searches.  The strongest constraint comes from the enhancement by a
factor of 9 of Higgs boson production by gluon fusion
\cite{Kribs:2007nz} due to the extra generation. Tevatron and ATLAS
have recently produced similar preliminary upper limits of $m_h < 124$
GeV \cite{Benjamin:2011sv} and $m_h < 120$ GeV \cite{atlas} at 95\%
c.l. through the decay channel $h\to W^+W^-$.  Our model avoids these
constraints to the extent that $\langle\tilde\phi\rangle = (g_B/\tilde g_B)
\cdot 232$ GeV can exceed the SM Higgs VEV.  Barring a large hierarchy
between these gauge couplings, 
it is unnatural to make  $\langle\tilde\phi\rangle$ very 
large, so one may anticipate a softening of this constraint rather
than completely evading it. In addition,  the new
Higgs boson $\phi$  can have large mixing with the SM Higgs, which
would lead to further softening of the constraint.

The model we have presented ties current hints for light dark matter
to light vector boson searches, Higgs physics, fourth generation
quarks, and the cosmology and
astrophysics of sterile (dark) neutrinos.  Even if this particular
model should be ruled out by future data, the idea of invisibly annihilating dark matter
may prove useful if its mass is at the 10 GeV scale. 

\medskip

{\bf Acknowledgments.}  We thank C.\ Burgess, R.\ Essig, J.\ Lavalle, J.\ Kamenik,
G.\ Raffelt, D.\ Spolyar, 
M.\ Trott, M.\ Williams, M.\ Wise, J.\ Zupan, for valuable discussions or 
correspondence.  JC thanks the Aspen Center for Physics for a
stimulating environment while this work was being done.

\end{document}